\documentclass[twocolumn,prb,showpacs,superscriptaddress]{revtex4}
\usepackage[T1]{fontenc}
\usepackage{color}
\usepackage{pifont}
\usepackage{textcomp}
\usepackage{amsmath}
\usepackage{amssymb}
\usepackage{graphicx}
\usepackage{esint}
\usepackage[unicode=true,pdfusetitle,
 bookmarks=true,bookmarksnumbered=false,bookmarksopen=false,
 breaklinks=false,pdfborder={0 0 1},backref=false,colorlinks=true]
 {hyperref}
\hypersetup{
 citecolor=blue}
\usepackage{breakurl}

\makeatletter
\@ifundefined{textcolor}{}
{%
 \definecolor{BLACK}{gray}{0}
 \definecolor{WHITE}{gray}{1}
 \definecolor{RED}{rgb}{1,0,0}
 \definecolor{GREEN}{rgb}{0,1,0}
 \definecolor{BLUE}{rgb}{0,0,1}
 \definecolor{CYAN}{cmyk}{1,0,0,0}
 \definecolor{MAGENTA}{cmyk}{0,1,0,0}
 \definecolor{YELLOW}{cmyk}{0,0,1,0}
}

\AtBeginDocument{
  
}

\makeatother

\begin{document}

\title{Probing zero-modes of defect in Kitaev quantum wire}

\author{Sheng-Wen Li}

\affiliation{Beijing Computational Science Research Center, Beijing 100084, China}

\affiliation{School of Engineering and Information Technology, University of New
South Wales at the Australian Defense Force Academy, Canberra, ACT
2600, Australia}

\author{Zeng-Zhao Li}

\affiliation{Beijing Computational Science Research Center, Beijing 100084, China}

\author{C. Y. Cai}

\affiliation{State Key Laboratory of Theoretical Physics, Institute of Theoretical
Physics, University of Chinese Academy of Sciences, Beijing 100190,
China}

\author{C. P. Sun}

\affiliation{Beijing Computational Science Research Center, Beijing 100084, China}
\begin{abstract}
The Kitaev quantum wire (KQW) model with open boundary possesses two
Majorana edge modes. When the local chemical potential on a defect
site is much higher than that on other sites and than the hopping
energy, the electron hopping is blocked at this site. We show that
the existence of such a defect on a closed KQW also gives rise to
two low-energy modes, which can simulate the edge modes. The energies
of the defect modes vanish to zero as the local chemical potential
of the defect increase to infinity. We develop a quantum Langevin
equation to study the transport of KQW for both open and closed cases.
We find that when the lead is contacted with the site beside the defect,
we can observe two splitted peaks around the zero-bias voltage in
the differential conductance spectrum. While if the lead is contacted
with the bulk of the quantum wire far from the the defect or the open
edges, we cannot observe any zero-bias peak.
\end{abstract}


\maketitle

\section{Introduction}

The emergent Majorana fermion in condensed matter system has attracted
much attention for its novel non-Abelian statistical property and
potential application in topological quantum computation \cite{kitaev_anyons_2006,kitaev_fault-tolerant_2003,nayak_non-abelian_2008}.
The Kitaev quantum wire (KQW) model with open boundary possesses two
localized Majorana edge modes at the two ends \cite{kitaev_unpaired_2001}.
The realization of this quantum wire model was also reported with
the help of strong spin-orbit coupling and Zeeman field in proximity
to an $s$-wave superconductor \cite{mourik_signatures_2012,das_zero-bias_2012,deng_anomalous_2012,churchill_superconductor-nanowire_2013,finck_anomalous_2013}.

To detect the existence of the Majorana fermion in the quantum wire,
people can measure the differential conductance in transport experiments
\cite{bolech_observing_2007,law_majorana_2009,flensberg_tunneling_2010,sau_non-abelian_2010}.
It was predicted that there is a zero-bias peak (ZBP) in the $dI/dV$
profile in the topological phase when the quantum wire is contacted
with a normal lead, and the height of the peak is $2e^{2}/h$ at zero
temperature. Moreover, if there exists finite coupling between the
Majorana fermions at the two ends, this ZBP would split into two peaks
\cite{flensberg_tunneling_2010}. However, it was recognized that
such feature of a single ZBP is not an unambiguous evidence, because
similar ZBP may be also induced by different mechanisms, such as Kondo
effect \cite{sasaki_kondo_2000,lee_zero-bias_2012,liu_zero-bias_2012,rainis_towards_2013,lee_probing_2013}.

When the local chemical potential $\mu_{p}$ on a defect site is much
higher than that on other sites and the hopping energy, the electron
hopping is blocked at this site. Thus, the existence of such a defect
on a closed wire is similar to cutting off the wire at this position
and generating new boundaries. Such ``cut off'' for a closed KQW
also gives rise to a pair of low-energy modes (we call them the \emph{defect
modes}). These defect modes have many similar properties to the Majorana
edge modes:
\begin{enumerate}
\item when the defect becomes ``strong'', i.e., the chemical potential
$\mu_{p}$ of the defect becomes quite large, the energies of the
defect modes approach zero;
\item the energies of the defect modes are gapped from the bulk band of
the quantum wire;
\item the defect modes are superpositions of both electron and hole modes
with equal weight, localized around the defect site.
\end{enumerate}
If $\mu_{p}$ approaches infinity, electron hopping is fully blocked
and the quantum wire can be regarded as completely cut off, thus the
defect modes become Majorana edge modes. In this sense, a close quantum
wire with a defect is equivalent to a homogenous open wire.

However, in practice, the strength of the defect is finite, thus there
remains a small energy splitting between the two defect modes. In
contrast, the energy splitting of Majorana edge modes in an open wire
is practically too small to be observed even for a short chain \cite{roy_majorana_2012}.
Throughout this paper, we call both the edge and defect modes the
\emph{zero-modes}.

With the above understanding, in this paper, we study the transport
measurement in KQW for two kinds of configurations, i.e., a homogenous
open wire and a closed wire with a defect. The detection of such zero-modes
induced by defect also helps test the existence of Majorana fermions
\cite{liu_andreev_2012,fregoso_electrical_2013}. We derive a quantum
Langevin equation to describe the electrical current transport of
the quantum wire contacted with two normal leads, which could give
exact numerical results \cite{dhar_quantum_2003,dhar_nonequilibrium_2006,yang_master_2013,roy_majorana_2012,roy_nature_2013}.
We obtain the differential conductance when one of the leads is contacted
with different sites of the quantum wire \cite{yang_proximity-effect-induced_2012}.
We find that, if the lead is contacted beside the defect, we can observe
two splitted ZBPs in the $dI/dV$ profile. Moreover, if the lead is
contacted with other sites in the bulk of the chain far from the defect
or the open edges, we cannot observe any ZBP, because the zero-modes,
both the edge and defect modes, are localized.

We arrange our paper as follows. In Sec.\,II we give a brief review
of KQW. We give a demonstration of the energy spectrum and spatial
distribution of the edge and defect modes. In Sec.\,III we derive
a quantum Langevin equation for the two contacts measurement setup,
and obtain the steady current formula. In Sec.\,IV, we show the $dI/dV$
profiles for different measurement configurations. Finally we draw
conclusion in Sec.\,V. We leave some mathematical tricks and details
of derivation in Appendix.

\section{Zero-modes of KQW}

\begin{figure}
\begin{centering}
\includegraphics[width=1\columnwidth]{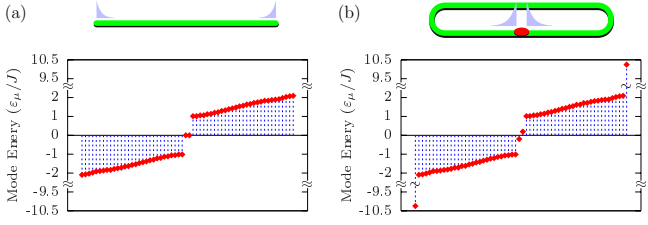}
\par\end{centering}

\caption{(Colored online) Demonstration of the energy spectrum of electrons
and holes for (a) homogenous open quantum wire (b) closed quantum
wire with a defect ($\mu_{p}=10$). The existence of a defect gives
rise to zero modes similar to open quantum wire. The chain has 30
sites, and other parameters are $\Delta=0.5,\,\mu=0.1$.}
\label{fig-spectrum}
\end{figure}

In this section, we present a brief review on KQW \cite{kitaev_unpaired_2001},
mainly in terms of fermion operators with respect to the normal modes
\cite{leijnse_introduction_2012}. We present a basic analysis on
the spectrum of the energy modes of the quantum wire system. We show
that the existence of a defect also gives rise to localized zero-modes,
which are similar to the Majorana edge modes under open boundary condition.

The KQW is a 1-dimensional tight-binding model plus a nearest pairing
term \cite{kitaev_unpaired_2001}. The Hamiltonian of the quantum
wire can be written as 
\begin{align}
\hat{H}_{\mathrm{w}}= & \sum_{i}J(\hat{d}_{i}^{\dagger}\hat{d}_{i+1}+\hat{d}_{i+1}^{\dagger}\hat{d}_{i})-\mu\hat{d}_{i}^{\dagger}\hat{d}_{i}\nonumber \\
 & -(\Delta\hat{d}_{i}^{\dagger}\hat{d}_{i+1}^{\dagger}+\Delta^{*}\hat{d}_{i+1}\hat{d}_{i}).\label{eq:H-w}
\end{align}

We denote $\mathbf{d}=(\hat{d}_{1},\dots,\hat{d}_{N},\hat{d}_{1}^{\dagger},\dots,\hat{d}_{N}^{\dagger})^{T}$,
where $N$ is the total number of the sites, then we can rewrite $\hat{H}_{\mathrm{w}}$
in a compact matrix form,
\begin{equation}
\hat{H}_{\mathrm{w}}=\frac{1}{2}\mathbf{d}^{\dagger}\cdot\mathbf{H}\cdot\mathbf{d},\quad\mathbf{H}=\left[\begin{array}{cc}
\mathbf{h} & \mathbf{p}\\
\mathbf{p}^{\dagger} & -\mathbf{h}
\end{array}\right],\label{eq:H-matrix}
\end{equation}
where $\mathbf{h}$ and $\mathbf{p}$ are $N\times N$ matrices, and
we omit a constant energy shift here. For an open wire, we have
\begin{equation}
\mathbf{h}=\mathbf{\left[\begin{array}{cccc}
-\mu & J\\
J & -\mu & \ddots\\
 & \ddots & \ddots & J\\
 &  & J & -\mu
\end{array}\right]},\,\mathbf{p}=\left[\begin{array}{cccc}
0 & -\Delta\\
\Delta & 0 & \ddots\\
 & \ddots & \ddots & -\Delta\\
 &  & \Delta & 0
\end{array}\right].\label{eq:hp}
\end{equation}
 While for periodic boundary condition, we should add $\mathbf{h}_{1,N}=\mathbf{h}_{N,1}=J$
and $\mathbf{p}_{1,N}=-\mathbf{p}_{N,1}=-\Delta$ to the above matrices.

The eigen modes of $\hat{H}_{\mathrm{w}}$ can be obtained by diagonalizing
the matrix $\mathbf{H}$. From $\mathbf{p}^{\dagger}=-\mathbf{p}^{*},\,\mathbf{h}^{\dagger}=\mathbf{h}$,
we can find that $\mathbf{H}$ has the following property \cite{blaizot_quantum_1986}:

\emph{Proposition}: If $\varepsilon$ is one eigenvalue of $\mathbf{H}$
with $\vec{V}=(v_{1},\dots,v_{N},w_{1},\dots,w_{N})^{T}$ as the eigenvector,
then $-\varepsilon$ is also an eigenvalue, and the corresponding
eigenvector is $\vec{V}'=(v_{1}^{*},\dots,v_{N}^{*},w_{1}^{*},\dots w_{N}^{*})^{T}$.

We also present a simple proof in Appendix A. This property roots
from the particle-hole symmetry, and guarantees that the eigen modes
of $\hat{H}_{\mathrm{w}}$ appear as particle-hole pairs,
\begin{align}
\hat{\psi}_{\mu} & =\sum_{i}\varphi^{\mu}{}_{i}\hat{d}_{i}+\phi^{\mu}{}_{i}\hat{d}_{i}^{\dagger},\nonumber \\
\hat{\psi}_{\mu}^{\dagger} & =\sum_{i}(\phi^{\mu}{}_{i})^{*}\hat{d}_{i}+(\varphi^{\mu}{}_{i})^{*}\hat{d}_{i}^{\dagger}:=\hat{\psi}_{\mu}',\label{eq:spatial}
\end{align}
 where $\hat{\psi}_{\mu}':=\hat{\psi}_{\mu}^{\dagger}$ can be regarded
as the modes for holes. But keep in mind that $\{\hat{\psi}_{\mu}',\,\hat{\psi}_{\mu}\}\neq0$
thus $\hat{\psi}_{\mu}'$ and $\hat{\psi}_{\mu}$ are not independent
fermion modes. Therefore we can always diagonalize the Hamiltonian
into the following form, 
\begin{equation}
\hat{H}_{\mathrm{w}}=\frac{1}{2}\mathbf{d}^{\dagger}\cdot\mathbf{H}\cdot\mathbf{d}=\frac{1}{2}\sum_{\mu=1}^{N}\varepsilon_{\mu}\hat{\psi}_{\mu}^{\dagger}\hat{\psi}_{\mu}-\varepsilon_{\mu}\hat{\psi}_{\mu}\hat{\psi}_{\mu}^{\dagger}.
\end{equation}

\begin{figure}
\begin{centering}
\includegraphics[width=0.8\columnwidth]{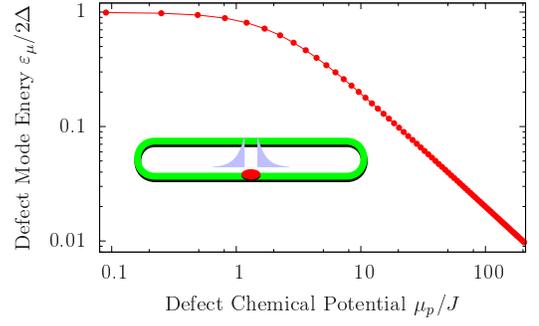}
\par\end{centering}

\caption{(Colored online) Mode energy of the defect mode in a closed wire ($N=30$).
We set $J=1,\,\Delta=0.6,\,\mu=0.1$. The energy of defect mode decreases
when $\mu_{p}$ increases.}
\label{fig-defect}
\end{figure}

A homogeneous open wire possesses two localized edge modes with zero
energy in the topological phase area $\left|\mu\right|<2\left|J\right|$,
but a homogeneous closed wire does not have such zero modes \cite{kitaev_unpaired_2001}.
If the local chemical potential $\mu_{p}$ on a defect site (site-$p$)
is much larger than that on other sites and than the hopping energy
$J$, the electron hopping is blocked at this site. Thus a closed
wire with a defect is similar to an open wire.

We demonstrate the energy spectrum of the quantum wire for both closed
and open configurations in Fig.\,\ref{fig-spectrum}. The existence
of a defect in a closed wire gives rise to two defect modes in the
superconducting gap separated from the bulk band. We also see that
there are two high-energy modes with $\varepsilon_{\mu}\simeq\pm\mu_{p}$
as the byproduct which we do not concern in this paper. Moreover,
the energies of the defect modes approach to zero when $\mu_{p}$
tends to infinity (Fig.\,\ref{fig-defect}).

\begin{figure}
\begin{centering}
\includegraphics[width=0.98\columnwidth]{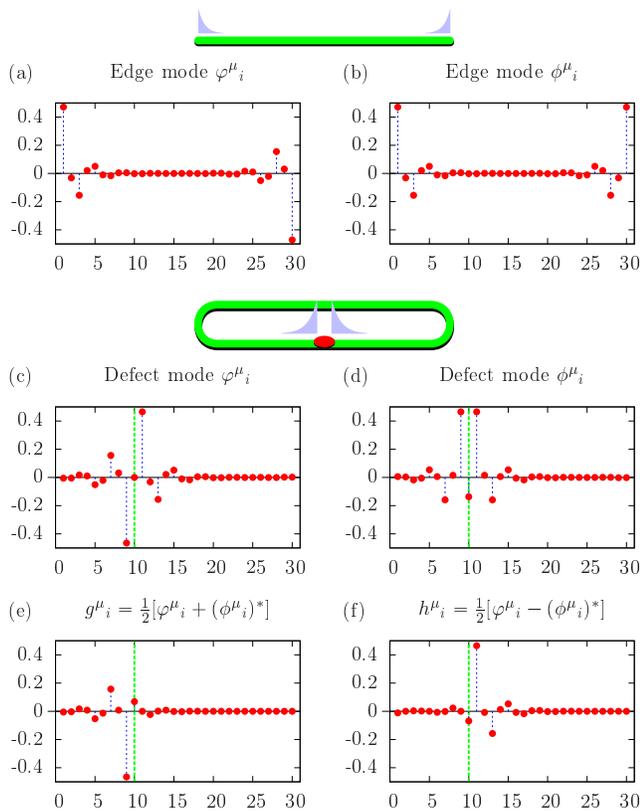}
\par\end{centering}

\caption{(Colored online) Spatial profile {[}$\varphi^{\mu}{}_{i}$ and $\phi^{\mu}{}_{i}$
in Eq.\,(\ref{eq:spatial}){]} of (a, b) edge mode in homogenous
open wire (c, d) defect mode in closed wire. (e, f) show the spatial
profile of the Majorana operator $\hat{\gamma}_{\mu,\pm}$ {[}$g^{\mu}{}_{i}$
and $h^{\mu}{}_{i}$ in Eq.\,(\ref{eq:MF-rep}){]}. The chain has
30 sites, and we set $J=1,\,\Delta=0.5,\,\mu=0.1$. There is a defect
with $\mu_{p}=10$ at the 10-th site (represented by the green vertical
line).}
\label{fig-profile}
\end{figure}

We also show the spatial distributions of the edge and defect modes
in Fig.\,\ref{fig-profile}(a-d) {[}namely, the coefficients $\varphi^{\mu}{}_{i}$
and $\phi^{\mu}{}_{i}$ in Eq.\,(\ref{eq:spatial}){]}. If we regard
the defect site as a new ``boundary'' of the quantum wire {[}dashed
line in Fig.\,\ref{fig-profile}(c, d){]}, we see that the spatial
distribution shapes of the defect modes in a closed quantum wire are
almost the same with that of the Majorana edge modes in an open wire
shown in Fig.\,\ref{fig-profile}(a, b). Therefore, in this sense,
a close quantum wire with a strong defect is equivalent to a homogenous
open wire.

For these localized zero-modes induced by defect, we can also represent
them by the Majorana operators \cite{kitaev_unpaired_2001},
\begin{align}
\hat{\gamma}_{\mu,+} & :=\hat{\psi}_{\mu}+\hat{\psi}_{\mu}^{\dagger}=\sum_{i}[g^{\mu}{}_{i}\hat{d}_{i}+(g^{\mu}{}_{i})^{*}\hat{d}_{i}^{\dagger}],\nonumber \\
\hat{\gamma}_{\mu,-} & :=-i(\hat{\psi}_{\mu}-\hat{\psi}_{\mu}^{\dagger})=-i\sum_{i}[h^{\mu}{}_{i}\hat{d}_{i}-(h^{\mu}{}_{i})^{*}\hat{d}_{i}^{\dagger}].\label{eq:MF-rep}
\end{align}
We show the spatial distributions of $\hat{\gamma}_{\mu,\pm}$ for
the defect modes in Fig.\,\ref{fig-profile}(e, f) {[}$g^{\mu}{}_{i}$
and $h^{\mu}{}_{i}$ in Eq.(\ref{eq:MF-rep}){]}. They are also Majorana
fermions, which are the anti-particles of themselves, i.e., $\hat{\gamma}_{\mu,\pm}=[\hat{\gamma}_{\mu,\pm}]^{\dagger}$.
With these notations, the effective Hamiltonian for the low-energy
modes can be written as 
\begin{equation}
\hat{H}_{\mathrm{low}}=\frac{i}{2}\sum_{\mu}\varepsilon_{\mu}\hat{\gamma}_{\mu,+}\hat{\gamma}_{\mu,-},\label{eq:H-eff}
\end{equation}
 where the summation includes the low-energy edge or defect modes.
For an open wire with finite length, or $\left|\mu_{p}/J\right|$
is not too large, $\varepsilon_{\mu}$ does not equal to zero exactly,
and Eq.\,(\ref{eq:H-eff}) is often regarded as the coupling between
the Majorana fermions \cite{kitaev_unpaired_2001,flensberg_tunneling_2010}.

Practically, for a homogenous open wire, the energy splitting of two
edge modes decays so fast with the length of the wire that we cannot
observe this splitting even for a quite short chain \cite{roy_majorana_2012}.
While for the defect modes, $\mu_{p}$ must be quite large ($\left|\mu_{p}/J\right|\gg100$)
so as to make $\varepsilon_{\mu}\simeq0$ (Fig.\,\ref{fig-defect}).
This property can be utilized to observe the splitting of the ZBP
in the $dI/dV$ spectrum.

\section{Quantum Langevin equation and steady current}

In this section, we derive a quantum Langevin equation to study the
transport behavior of KQW contacted with two electron leads. And we
obtain the formula for the steady current.

\subsection{Quantum Langevin equation}

\begin{figure}
\begin{centering}
\includegraphics[width=1\columnwidth]{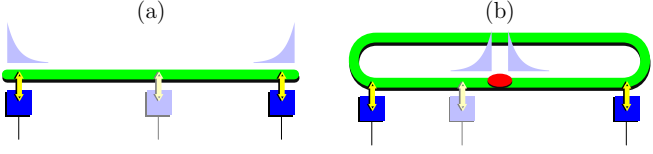}
\par\end{centering}

\caption{(Colored online) Two contacts transport measurement setup for (a)
homogenous open wire (b) closed wire with a defect. The leads can
be contacted with different sites.}

\label{fig-measure}
\end{figure}

We derive a quantum Langevin equation to study the transport of this
quantum wire \cite{dhar_quantum_2003,roy_majorana_2012,roy_nature_2013}.
The transport measurement setup of the quantum wire is demonstrated
in Fig.\,\ref{fig-measure}. Our derivation here is valid for both
the open and closed quantum wire cases. We consider the quantum wire
of $N$ sites coupled with two normal leads via electron tunneling
at site-$x,y$ ($1\le x,y\le N$) respectively. The total Hamiltonian
of the quantum wire and the leads can be written as
\begin{equation}
{\cal H}=\hat{H}_{\mathrm{w}}+\hat{H}_{B}+\hat{H}_{T},
\end{equation}
where $\hat{H}_{\mathrm{w}}$ is shown in Eq.\,(\ref{eq:H-w}). $\hat{H}_{B}$
is the Hamiltonian for the two electron leads contacting with site-$x,y$,
\begin{equation}
\hat{H}_{B}=\sum_{\mathbf{k}_{x}}\omega_{\mathbf{k}_{x}}\hat{c}_{\mathbf{k}_{x}}^{\dagger}\hat{c}_{\mathbf{k}_{x}}+\sum_{\mathbf{k}_{y}}\omega_{\mathbf{k}_{y}}\hat{c}_{\mathbf{k}_{y}}^{\dagger}\hat{c}_{\mathbf{k}_{y}}.
\end{equation}
 $H_{T}$ describes the tunneling between the quantum wire and the
leads, 
\begin{equation}
\hat{H}_{T}=\hat{d}_{x}^{\dagger}\hat{\Psi}_{x}+\hat{\Psi}_{x}^{\dagger}\hat{d}_{x}+\hat{d}_{y}^{\dagger}\hat{\Psi}_{y}+\hat{\Psi}_{y}^{\dagger}\hat{d}_{y}
\end{equation}
where $\hat{\Psi}_{x}=\sum_{\mathbf{k}_{x}}g_{\mathbf{k}_{x}}\hat{c}_{\mathbf{k}_{x}}$.

We assume that the system evolves from $t=0$, i.e., $\hat{d}_{i}(t)=\Theta(t)\hat{d}_{i}(t)$
and $\hat{c}_{\mathbf{k}_{x}}(t)=\Theta(t)\hat{c}_{\mathbf{k}_{x}}(t)$,
and at the initial time, each lead stays at a canonical state, $\rho_{x}\propto\exp[-\beta_{x}\sum(\omega_{\mathbf{k}_{x}}-\mu_{x})\hat{c}_{\mathbf{k}_{x}}^{\dagger}\hat{c}_{\mathbf{k}_{x}}]$,
where $\mu_{x}$ is the chemical potential and $\beta_{x}^{-1}=T_{x}$
is the temperature for lead-$x$.

We start from the Heisenberg equation \cite{yang_master_2013},
\begin{equation}
\partial_{t}[\Theta(t)\hat{O}(t)]=\delta(t)\hat{O}(0)-i\Theta(t)[\hat{O}(t),\,{\cal H}],
\end{equation}
where $\hat{O}(t)$ may be the operator $\hat{c}_{\mathbf{k}_{x}}(t)$
or $\hat{d}_{i}(t)$. The equations of motion for $\hat{c}_{\mathbf{k}_{x}}(t)$
and $\hat{d}_{x}(t)$ are
\begin{align}
\partial_{t}\hat{c}_{\mathbf{k}_{x}}(t) & =\delta(t)\hat{c}_{\mathbf{k}_{x}}(0)-i(\omega_{\mathbf{k}_{x}}\hat{c}_{\mathbf{k}_{x}}+g_{\mathbf{k}_{x}}^{*}\hat{d}_{x}),\nonumber \\
\partial_{t}\hat{d}_{x}(t) & =\delta(t)\hat{d}_{x}(0)-i[\hat{H}_{\mathrm{w}},\hat{d}_{x}]-i\sum_{\mathbf{k}_{x}}g_{\mathbf{k}_{x}}\hat{c}_{\mathbf{k}_{x}}.
\end{align}

We integrate the equation of $\hat{c}_{\mathbf{k}_{x}}(t)$ and obtain
\begin{align}
\hat{c}_{\mathbf{k}_{x}}(t)= & \Theta(t)\hat{c}_{\mathbf{k}_{x}}(0)e^{-i\omega_{\mathbf{k}_{x}}t}\nonumber \\
 & \quad-ig_{\mathbf{k}_{x}}^{*}\int_{0}^{t}d\tau\, e^{-i\omega_{\mathbf{k}_{x}}(t-\tau)}\hat{d}_{x}(\tau).\label{eq:c(t)}
\end{align}
Inserting it into the equation of $\hat{d}_{x}(t)$ above, we obtain
a differential-integral equation,
\begin{align}
\partial_{t}\hat{d}_{x}= & \delta(t)\hat{d}_{x}(0)-i[\hat{H}_{\mathrm{w}},\hat{d}_{x}]\nonumber \\
 & \quad-i\hat{\eta}_{x}(t)-\int_{0}^{t}d\tau\, D_{x}(t-\tau)\hat{d}_{x}(\tau),
\end{align}
where 
\begin{align}
\hat{\eta}_{x}(t) & :=\Theta(t)\sum_{\mathbf{k}_{x}}g_{\mathbf{k}_{x}}\hat{c}_{\mathbf{k}_{x}}(0)e^{-i\omega_{\mathbf{k}_{x}}t},\nonumber \\
D_{x}(t) & :=\Theta(t)\sum_{\mathbf{k}_{x}}\left|g_{\mathbf{k}_{x}}\right|^{2}e^{-i\omega_{\mathbf{k}_{x}}t}.\label{eq:randomForce}
\end{align}
$\hat{\eta}_{x}(t)$ is the random force and $D_{x}(t)$ is the damping
kernel. These dissipation terms do not appear in the equations of
$\hat{d}_{i}(t)$ for $i\neq x,y$.

We can write down the quantum Langevin equation for $\mathbf{d}=(\hat{d}_{1},\dots,\hat{d}_{N},\hat{d}_{1}^{\dagger},\dots,\hat{d}_{N}^{\dagger})^{T}$
in a compact matrix form, 
\begin{align}
\partial_{t}\mathbf{d}= & \delta(t)\mathbf{d}(0)-i\mathbf{H}\cdot\mathbf{d}\nonumber \\
 & \quad-i\boldsymbol{\eta}(t)-\int_{0}^{t}d\tau\,\mathbf{D}(t-\tau)\cdot\mathbf{d}(\tau).\label{eq:QLE}
\end{align}
Here $\mathbf{D}(t)=\mathbf{D}_{x}(t)+\mathbf{D}_{y}(t)$ is a diagonalized
$2N\times2N$ matrix, while $\boldsymbol{\eta}(t)=\boldsymbol{\eta}^{x}(t)+\boldsymbol{\eta}^{y}(t)$
is a vector of $2N$-dimension. The elements of the damping matrix
$\mathbf{D}^{x}(t)$ are 
\begin{equation}
[\mathbf{D}^{x}(t)]_{ij}=\begin{cases}
D_{x}(t), & i=j=x,\\
D_{x}^{*}(t), & i=j=N+x,\\
0, & \text{others}.
\end{cases}
\end{equation}
While the elements of the random force vector $\boldsymbol{\eta}^{x}(t)$
are
\begin{equation}
[\boldsymbol{\eta}^{x}(t)]_{i}=\begin{cases}
\hat{\eta}_{x}(t), & i=x,\\
-\hat{\eta}_{x}(t), & i=N+x,\\
0, & \text{others}.
\end{cases}
\end{equation}
 We should also notice that the integral limit in Eq.\,(\ref{eq:QLE})
can be extended to $\pm\infty$ since we have $\mathbf{d}(t)=\Theta(t)\mathbf{d}(t)$
and $\mathbf{D}(t)=\Theta(t)\mathbf{D}(t)$. Our derivation here is
valid for both the open and closed wire cases. For different quantum
wire configurations, we just need to change the matrix $\mathbf{H}$
{[}Eq.\,(\ref{eq:H-matrix}){]}.

\subsection{Steady current formula}

Formally, the above quantum Langevin equation (\ref{eq:QLE}) can
be solved exactly by Fourier transform
\[
\tilde{f}(\omega)=\int_{-\infty}^{\infty}dt\,\hat{f}(t)e^{i\omega t},\quad\hat{f}(t)=\int_{-\infty}^{\infty}\frac{d\omega}{2\pi}\,\tilde{f}(\omega)e^{-i\omega t}.
\]
The Fourier transform of Eq.\,(\ref{eq:QLE}) gives
\[
-i\omega\tilde{\mathbf{d}}(\omega)=\mathbf{d}(0)-i\mathbf{H}\cdot\tilde{\mathbf{d}}(\omega)-i\tilde{\boldsymbol{\eta}}(\omega)-\tilde{\mathbf{D}}(\omega)\cdot\tilde{\mathbf{d}}(\omega).
\]
 Thus we have
\begin{align}
\tilde{\mathbf{d}}(\omega) & =\tilde{\mathbf{G}}(\omega)[\mathbf{d}(0)-i\tilde{\boldsymbol{\eta}}(\omega)],\nonumber \\
\tilde{\mathbf{G}}(\omega) & =i[\omega-\mathbf{H}+i\tilde{\mathbf{D}}(\omega)]^{-1},\label{eq:solution}
\end{align}
 where $\tilde{\mathbf{G}}(\omega)$ is the propagator matrix.

Here we introduce the coupling spectrum $\Gamma_{x}(\omega):=2\pi\sum_{\mathbf{k}_{x}}$$\left|g_{\mathbf{k}_{x}}\right|^{2}\delta(\omega-\omega_{\mathbf{k}_{x}})$,
and then the damping kernel $D_{x}(t)$ can be rewritten as
\begin{align}
D_{x}(t) & =\Theta(t)\int_{-\infty}^{\infty}\frac{d\omega}{2\pi}\,\Gamma_{x}(\omega)e^{-i\omega t},\nonumber \\
\tilde{D}_{x}(\omega) & =\frac{1}{2}\Gamma_{x}(\omega)+i\mathbf{P}\int\frac{d\nu}{2\pi}\,\frac{\Gamma_{x}(\nu)}{\omega-\nu}.
\end{align}
 The real part of $\tilde{D}_{x}(\omega)$ describes the dissipation
while the imaginary part is the self-energy correction. We denote
$\tilde{\boldsymbol{\Gamma}}(\omega):=\tilde{\mathbf{D}}+\tilde{\mathbf{D}}^{\dagger}$
as the dissipation matrix, which is the real part of $2\tilde{\mathbf{D}}(\omega)$.
Once the coupling spectrums of the two leads are given, in principle
we can obtain the propagator matrix $\tilde{\mathbf{G}}(\omega)$
and the dynamics of $\mathbf{d}(t)$ exactly. Here we take the spectrum
to be a Lorentzian function,
\begin{equation}
\Gamma_{x}(\omega)=\Gamma_{y}(\omega)=\frac{\lambda\Omega_{c}^{2}}{\omega^{2}+\Omega_{c}^{2}},
\end{equation}
where $\Omega_{c}$ is the cutoff frequency and $\lambda$ describes
the tunneling strength with the quantum wire. With this spectrum,
the self-energy correction is zero.

Now we can derive the steady current when $t\rightarrow\infty$. The
electrical current flowing out of the lead contacted with site-$x$
can be defined from the changing rate of the electron number in this
lead \cite{flensberg_tunneling_2010,liu_andreev_2012}, 
\begin{equation}
\hat{I}_{x}(t)=-\frac{ie}{\hbar}[\hat{d}_{x}^{\dagger}(t)\hat{\Psi}_{x}(t)-\hat{\Psi}_{x}^{\dagger}(t)\hat{d}_{x}(t)].
\end{equation}

In this open quantum system, the current $\overline{I}(t)$ would
approach a steady state after a long time evolution. This steady current
can be obtained from the Fourier transform $\tilde{I}_{x}(\omega)$
of $\langle\hat{I}_{x}(t)\rangle$ (see Appendix \ref{sec:Steady-current-formula}),
\begin{align}
\overline{I}_{x}(t\rightarrow\infty)= & -i\lim_{\omega\rightarrow0}\big[\omega\tilde{I}(\omega)\big],\nonumber \\
\tilde{I}_{x}(\omega)= & -\frac{ie}{\hbar}\int\frac{d\nu}{2\pi}\langle\tilde{d}_{x}^{\dagger}(\nu)\tilde{\Psi}_{x}(\nu+\omega)\rangle\nonumber \\
 & \qquad-\langle\tilde{\Psi}_{x}^{\dagger}(\nu+\omega)\tilde{d}_{x}(\nu)\rangle.\label{eq:steady-current}
\end{align}
Recall that $\hat{\Psi}_{x}(t)=\sum_{\mathbf{k}_{x}}g_{\mathbf{k}_{x}}\hat{c}_{\mathbf{k}_{x}}(t)$,
combining with Eqs.\,(\ref{eq:c(t)}, \ref{eq:randomForce}), we
have 
\begin{equation}
\tilde{\Psi}_{x}(\omega)=\tilde{\eta}_{x}(\omega)-i\tilde{D}_{x}(\omega)\tilde{d}_{x}(\omega).
\end{equation}
With the help of the solution of $\tilde{d}_{x}(\omega)$ {[}Eq.\,(\ref{eq:solution}){]},
all the expectation in $\tilde{I}_{x}(\omega)$ {[}Eq.\,(\ref{eq:steady-current}){]}
can be expressed by the fluctuation of the random forces $\langle\tilde{\eta}_{x}^{\dagger}(\nu+\omega)\tilde{\eta}_{y}(\nu)\rangle$,
which relates to the coupling spectrum $\Gamma_{x,y}(\omega)$ and
the Fermi distribution $f_{x,y}(\omega-\mu_{x,y})$ of each lead.
We obtain the formula for the steady current below (see derivation
in Appendix \ref{sec:Fluctuation-relation}),
\begin{align}
\overline{I}_{x}(t\rightarrow\infty)= & \frac{e}{\hbar}\int\frac{d\omega}{2\pi}\tilde{\mathbf{G}}_{yx}^{\dagger}\Gamma_{x}\tilde{\mathbf{G}}_{xy}\Gamma_{y}(f_{x}-f_{y})\nonumber \\
 & +\tilde{\mathbf{G}}_{y+N,x}^{\dagger}\Gamma_{y}\tilde{\mathbf{G}}_{x,y+N}\Gamma_{x}(f_{x}-\overline{f}_{y})\nonumber \\
 & +\tilde{\mathbf{G}}_{x+N,x}^{\dagger}\Gamma_{x}\tilde{\mathbf{G}}_{x,x+N}\Gamma_{x}(f_{x}-\overline{f}_{x}),\label{eq:current}
\end{align}
where we denote $f_{x}:=f_{x}(\omega-\mu_{x})$ and $\overline{f}_{x}:=f_{x}(\mu_{x}-\omega)=1-f_{x}$.
Here $f_{x}(\omega-\mu_{x}):=[\exp\beta_{x}(\omega-\mu_{x})+1]^{-1}$
is the Fermi distribution.

If there is no pairing terms in the quantum wire Hamiltonian Eq.\,(\ref{eq:H-w}),
$\tilde{\mathbf{G}}$ is block-diagonlized and the last two terms
in Eq.\,(\ref{eq:current}) do not appear. In this case this formula
returns to the result for a tight-binding chain \cite{meir_landauer_1992,datta_quantum_2005}.
On the other hand, the last term is similar to the current formula
derived in Ref.\,\cite{flensberg_tunneling_2010}, where only the
Majorana subspace is considered.

We can regard $\tilde{\mathbf{G}}_{ij}(\omega)$ as the transition
amplitude between different modes, and this current formula can be
understood in an intuitive picture. The first term in Eq.\,(\ref{eq:current})
represents the transition between the local modes $\hat{d}_{x}$ and
$\hat{d}_{y}$, i.e., an electron emits from lead-$x$, and then it
is received as an electron by lead-$y$. Since there is superconducting
pairing effect, the electron emitted from lead-$x$ can be also received
as a hole by lead-$y$, as represented by the second term. The last
term represents the transition between the electron and hole modes
both at site-$x$, and indeed this term gives the main contribution
to the ZBPs that come from the zero-modes.

\section{Differential conductance for KQW}

\begin{figure}
\begin{centering}
\includegraphics[width=1\columnwidth]{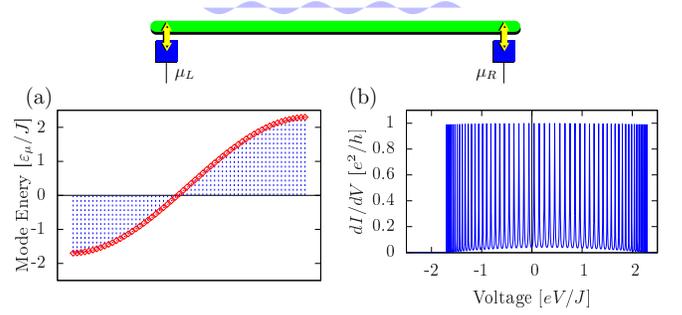}
\par\end{centering}

\caption{(Colored online) Differential conductance for a homogenous tight-binding
chain ($N=60$). We set $J=1,\,\mu=0.3,\,\Delta=0,\,\lambda=0.2,\,\Omega_{c}=20$.
(a) The energy spectrum. (b) The differential conductance $dI/dV$.
The leads contact with the two ends.}

\label{fig-TB}
\end{figure}

Now we have obtained the steady current. We set the chemical potential
of the left lead as $\mu_{x}=(-e)V$, while we keep $\mu_{y}=(-e)V_{0}$
as constant. At the zero temperature, $f_{x}(\omega-\mu_{x})=\Theta(\omega-\mu_{x})$,
thus we obtain the differential conductance as
\begin{align}
\frac{dI}{dV}= & \frac{e^{2}}{h}\big[\tilde{\mathbf{G}}_{yx}^{\dagger}\Gamma_{x}\tilde{\mathbf{G}}_{xy}\Gamma_{y}+\tilde{\mathbf{G}}_{y+N,x}^{\dagger}\Gamma_{y}\tilde{\mathbf{G}}_{x,y+N}\Gamma_{x}\nonumber \\
 & +2\tilde{\mathbf{G}}_{x+N,x}^{\dagger}\Gamma_{x}\tilde{\mathbf{G}}_{x,x+N}\Gamma_{x}\big](eV).\label{eq:dIdV}
\end{align}

The above differential conductance formula is exact. We can calculate
$\tilde{\mathbf{G}}_{ij}(\omega)$ numerically to get the $dI/dV$
spectrum. When there is no superconducting pairing term, $\Delta=0$,
the system becomes a tight-binding chain. The last two terms in Eq.\,(\ref{eq:dIdV})
all vanish to zero. The energy spectrum for electron modes (no hole
modes) and $dI/dV$ profile for an open tight-binding chain is shown
in Fig.\,\ref{fig-TB}. We see that for a homogenous tight-binding
chain contacted with two leads at each end, each mode in the conducting
band gives rise to a peak whose height is one conductance quantum
$G_{0}=e^{2}/h$, and the positions of the peaks correspond to the
mode energies \cite{bruus_many-body_2004,datta_quantum_2005}.

\begin{figure}
\begin{centering}
\includegraphics[width=1\columnwidth]{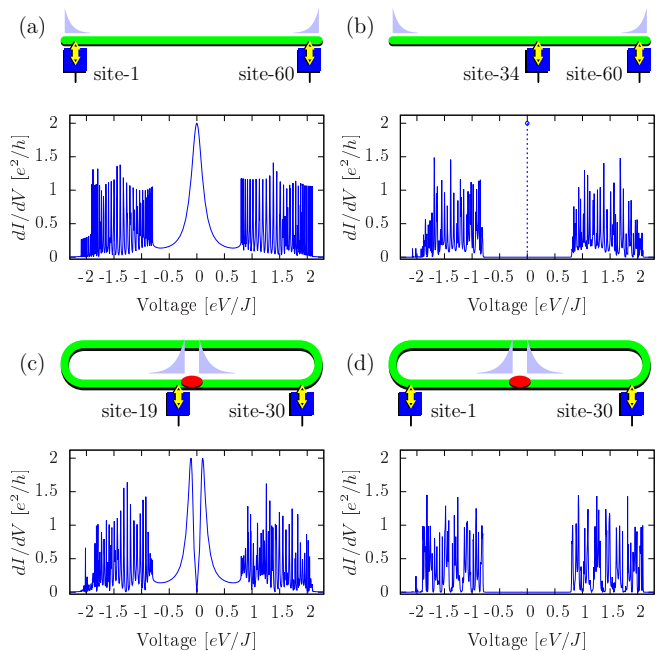}
\par\end{centering}

\caption{(Colored online) Differential conductance in topological phase regime
for (a, b) homogenous open wire (c,d) closed wire with a defect ($\mu_{p}=15$
at site-20). We set $J=1,\,\mu=0.1,\,\Delta=0.4,\,\lambda=0.3,\,\Omega_{c}=20$.
The chain contains 60 sites. The position of the right lead is fixed,
while the left lead is contacted with different sites.}

\label{fig-Topo}
\end{figure}

In Fig.\,\ref{fig-Topo}, we show the $dI/dV$ profile (from the
left lead) in the topological phase regime that affords zero-modes
for different measurement configurations. We fix the position of the
right lead, while the left lead can be contacted with different sites.
The peaks on the two sides are contributed from the band modes, corresponding
to the band structure shown in Fig.\,\ref{fig-spectrum}. 

For the homogenous open wire, when the lead is contacted with the
left edge {[}Fig.\,\ref{fig-Topo}(a){]}, there is a single ZBP which
is contributed from the the edge modes. The energy gap of the two
edge modes can be neglected for a chain long enough ($N=60$ here).
If the lead is contacted with a bulk site far from the edges, we cannot
observe any ZBP. Here our numerical result gives a singular point
at $V=0$. Precisely speaking, the width of this ZBP is zero, and
practically this peak is ``unobservable''. However, we still plot
out this singular point in the picture {[}dashed line in Fig.\,\ref{fig-Topo}(b){]}.

For the closed wire with a defect, when the lead is contacted with
the nearest site beside the defect {[}Fig.\,\ref{fig-Topo}(c){]},
we can observe two splitted peaks, whose positions correspond to the
energies of the defect modes. Their heights are also $2e^{2}/h$.
The distance between the two peaks depends on the strength of the
defect, or rather, the coupling between the Majorana fermions. While
if the lead is contacted with a bulk site far from the defect, again
we cannot observe any ZBP {[}Fig.\,\ref{fig-Topo}(d){]}.

Our results also indicate that whether a normal mode $\hat{\psi}_{\mu}$
{[}Eq.\,(\ref{eq:spatial}){]} can be observed in the $dI/dV$ spectrum
depends on its overlap with the local mode $\hat{d}_{x}$ contacted
with the lead.

\section{Conclusion}

In this paper, we studied the transport measurement in KQW for two
kinds of configurations, i.e., a homogenous open wire and a closed
wire with a defect. The existence of a defect also gives rise to a
pair of zero-modes, which are localized superpositions of both electron
and hole modes. The behavior of a defect is similar to two open edges.

We derived a quantum Langevin equation to study the two-contact transport.
We obtained the formulas for the steady electrical current and differential
conductance. We obtained the exact numerical result for the $dI/dV$
spectrum when one of the leads is contacted with different site of
the chain. When the lead is contacted with the edge of the open quantum
wire, we can observe a single ZBP with height $2e^{2}/h$ contributed
from the degenerated Majorana edge modes. While if the lead is contacted
with the site beside the defect in the closed wire, we can observe
two splitted ZBPs contributed from the defect modes. The heights of
the two peaks are also $2e^{2}/h$. When the lead is contacted with
other sites in the bulk far from the defect or the open edges, no
ZBP can be observed.

Base on the above results, we would suggest that a group of comparison
experiments for different transport measurement configurations, as
we have shown above, may be helpful to test the existence of Majorana
fermion.

\emph{Acknowledgement - }This work is supported by National Natural
Science Foundation of China under Grants Nos.\,11121403, 10935010
and 11074261, National 973-program Grants No.\,2012CB922104, and
Postdoctoral Science Foundation of China No.\,2013M530516. S.-W.
Li is grateful to University of New South Wales at the Australian
Defense Force Academy for its hospitality during his visit.

\appendix

\section{Property of the quantum wire Hamiltonian matrix}

Here we present the proof of the property of the Hamiltonian matrix
$\mathbf{H}$ {[}see Eq.\,(\ref{eq:H-matrix}){]} mentioned in Sec.\,II.

\emph{Proposition}: If $\varepsilon$ is one eigenvalue of $\mathbf{H}$
with $\vec{V}=(v_{1},\dots,v_{N},w_{1},\dots,w_{N})^{T}$ as the eigenvector,
then $-\varepsilon$ is also an eigenvalue, and the corresponding
eigenvector is $\vec{V}'=(w_{1}^{*},\dots,w_{N}^{*},w_{1}^{*},\dots w_{N}^{*})^{T}$.

\emph{Proof}: The eigen equation of $\mathbf{H}\vec{V}=\varepsilon\vec{V}$
is
\[
\left[\begin{array}{cc}
\mathbf{h} & \mathbf{p}\\
\mathbf{p}^{\dagger} & -\mathbf{h}
\end{array}\right]\left(\begin{array}{c}
\boldsymbol{v}\\
\boldsymbol{w}
\end{array}\right)=\varepsilon\left(\begin{array}{c}
\boldsymbol{v}\\
\boldsymbol{w}
\end{array}\right).
\]
 Or we can write it as
\begin{align*}
\mathbf{h}_{ij}v_{j}+\mathbf{p}_{ij}w_{j} & =\varepsilon v_{i},\\
-\mathbf{p}_{ij}^{*}v_{j}-\mathbf{h}_{ij}w_{j} & =\varepsilon w_{i}.
\end{align*}
From the explicit form of $\mathbf{h}$ and $\mathbf{p}$ {[}see Eq.\,(\ref{eq:hp}){]}
we should notice that $\mathbf{p}^{\dagger}=-\mathbf{p}^{*},\,\mathbf{h}^{\dagger}=\mathbf{h}$.
Thus, the negative conjugation of the above two equations gives
\begin{align*}
\mathbf{h}_{ij}w_{j}^{*}+\mathbf{p}_{ij}v_{j}^{*} & =-\varepsilon w_{i}^{*},\\
-\mathbf{p}_{ij}^{*}w_{j}^{*}-\mathbf{h}_{ij}v_{j}^{*} & =-\varepsilon v_{i}^{*}.
\end{align*}
 Or we can write it as 
\[
\left[\begin{array}{cc}
\mathbf{h} & \mathbf{p}\\
\mathbf{p}^{\dagger} & -\mathbf{h}
\end{array}\right]\left(\begin{array}{c}
\boldsymbol{w}^{*}\\
\boldsymbol{v}^{*}
\end{array}\right)=-\varepsilon\left(\begin{array}{c}
\boldsymbol{w}^{*}\\
\boldsymbol{v}^{*}
\end{array}\right).
\]
This is just the eigen equation $\mathbf{H}\vec{V}'=-\varepsilon\vec{V}'$.
$\qquad\blacksquare$

\section{General steady formula\label{sec:Steady-current-formula}}

We want to study the long time behavior of some dynamical observable,
e.g., the electrical current $I(t\rightarrow\infty)$. Here we have
a method to evaluate the long time behavior of $I(t)$ from the poles
of its Fourier transform $\tilde{I}(\omega)$.

By Fourier transform, we have
\begin{equation}
I(t)=\int\frac{d\omega}{2\pi}\,\tilde{I}(\omega)e^{-i\omega t}.\label{eq:I(t)}
\end{equation}
We should notice that if $I(t)$ diverges when $t\rightarrow\infty$,
indeed $\tilde{I}(\omega)$ does not exist. If $\tilde{I}(\left|\omega\right|\rightarrow\infty)\rightarrow0$,
Eq.\,(\ref{eq:I(t)}) can be integrated by the residue theorem. For
$t>0$, the contour integral takes the lower loop, and we have
\begin{equation}
I(t)=-i\sum_{\substack{\text{lower}\\
\text{plane}
}
}\mathrm{Res}[\tilde{I}(\omega)e^{-i\omega t}]-\frac{i}{2}\sum_{\substack{\text{real}\\
\text{axis}
}
}\mathrm{Res}[\tilde{I}(\omega)e^{-i\omega t}],\label{eq:I(t)-I(w)}
\end{equation}
 where the summations contain all the poles in the lower plane and
on the real axis respectively.

Consider the case that the pole at $\omega_{r}$ is simple, we have
\begin{align}
\mathrm{Res}_{\omega_{r}}[\tilde{I}(\omega)e^{-i\omega t}] & =\lim_{\omega\rightarrow\omega_{r}}\big[(\omega-\omega_{r})\tilde{I}(\omega)e^{-i\omega t}\big]\nonumber \\
 & =e^{-i\omega_{r}t}\mathrm{Res}_{\omega_{r}}[\tilde{I}(\omega)].
\end{align}
We see that all the time dependence of $I(t)$ is contained in $\exp[-i\omega_{r}t]$.
There are three types of poles here:

1) In the lower plane, $\omega_{r}=\omega_{0}-i\gamma$, and $\gamma>0$.
For these poles, $\exp[-i\omega_{r}t]$ gives rise to terms with exponential
decay behavior, and they vanish when $t\rightarrow+\infty$;

2) On the real axis, or infinitely close to it in the lower plane,
$\omega_{r}=\omega_{0}-i0^{+}$, but $\omega_{r}\neq0$. These poles
contribute to terms that keep oscillating at frequency $\omega_{0}$
when $t\rightarrow+\infty$;

3) At the origin point $\omega_{r}=0$. This pole contributes a time-independent
term.

Therefore, we can evaluate the long time behavior of $I(t)$ from
the poles of its Fourier transform $\tilde{I}(\omega)$. If $\tilde{I}(\omega)$
has no other poles near the real axis in the lower plane, except the
simple pole $\omega_{r}=0-i\epsilon^{+}$, we can write down the steady
state of $I(t)$ from Eq.\,(\ref{eq:I(t)-I(w)}) as
\begin{equation}
I(t\rightarrow+\infty)=-i\lim_{\omega\rightarrow0}\big[\omega\tilde{I}(\omega)\big].\label{eq:steady-formula}
\end{equation}

For example, we consider a current that decays exponentially from
$t_{0}=0$, 
\[
I(t)=\Theta(t)I_{0}(1-e^{-i\omega_{0}t-\gamma t}).
\]
The Fourier transform of $I(t)$ is 
\[
\tilde{I}(\omega)=\frac{i}{\omega+i\epsilon^{+}}+\frac{i}{\omega-\omega_{0}+i\gamma},
\]
 and we can check that Eq.\,(\ref{eq:steady-formula}) holds.

\section{Steady current\label{sec:Fluctuation-relation}}

First, we calculate the fluctuation relation of the random forces
$\langle\tilde{\eta}_{x}^{\dagger}(\omega)\tilde{\eta}_{y}(\omega')\rangle$.
The random force acting on the $x$-th contact site is 
\begin{align}
\hat{\eta}_{x}(t) & =\Theta(t)\sum_{\mathbf{k}_{x}}g_{\mathbf{k}_{x}}e^{-i\omega_{\mathbf{k}_{x}}t}\hat{c}_{\mathbf{k}_{x}}(0),\nonumber \\
\tilde{\eta}_{x}(\omega) & =\sum_{\mathbf{k}_{x}}\frac{ig_{\mathbf{k}_{x}}\hat{c}_{\mathbf{k}_{x}}(0)}{\omega-\omega_{\mathbf{k}_{x}}+i\epsilon^{+}},
\end{align}
where $\tilde{\eta}_{x}(\omega)$ is the Fourier transform of $\hat{\eta}_{x}(t)$.

We have assumed that initially each reservoir stays at a canonical
thermal state, which gives
\begin{align}
\langle\hat{c}_{\mathbf{k}_{x}}^{\dagger}(0)\hat{c}_{\mathbf{q}_{y}}(0)\rangle= & \delta_{xy}f_{x}(\omega_{\mathbf{k}_{x}}-\mu_{x}),\nonumber \\
f_{x}(\omega-\mu):= & [\exp\frac{\omega-\mu}{kT_{x}}+1]^{-1}.
\end{align}
Here $f_{x}(\omega-\mu)$ is the Fermi distribution, and $\mu$ is
the chemical potential. Thus, we have
\begin{align}
\langle\tilde{\eta}_{x}^{\dagger}(\omega)\tilde{\eta}_{y}(\omega')\rangle & =\sum_{\mathbf{k}_{x},\mathbf{q}_{y}}\frac{g_{\mathbf{k}_{x}}^{*}g_{\mathbf{q}_{y}}\langle\hat{c}_{\mathbf{k}_{x}}^{\dagger}(0)\hat{c}_{\mathbf{q}_{y}}(0)\rangle}{(\omega-\omega_{\mathbf{k}_{x}}-i\epsilon^{+})(\omega'-\omega_{\mathbf{q}_{y}}+i\epsilon^{+})}\nonumber \\
 & =\int\frac{d\nu}{2\pi}\,\frac{\Gamma_{x}(\nu)f_{x}(\nu-\mu_{x})\delta_{xy}}{(\omega-\nu-i\epsilon^{+})(\omega'-\nu+i\epsilon^{+})}\nonumber \\
 & =\frac{i\Gamma_{x}(\omega')f_{x}(\omega'-\mu_{x})\delta_{xy}}{(\omega'-\omega)+2i\epsilon^{+}},\nonumber \\
\langle\tilde{\eta}_{x}(\omega)\tilde{\eta}_{y}^{\dagger}(\omega')\rangle & =\frac{i\Gamma_{x}(\omega')f_{x}(\mu_{x}-\omega')\delta_{xy}}{(\omega'-\omega)+2i\epsilon^{+}},\label{eq:fluc}
\end{align}
 where we should notice that $1-f(\omega-\mu)=f(\mu-\omega)$. The
above integrals are done by residue theorem.

From Appendix\,\ref{sec:Steady-current-formula} and Eq.\,(\ref{eq:steady-current})
we see that the calculation of the steady current $\overline{I}(t\rightarrow\infty)$
requires us to evaluate expectation values of the following form,
\begin{equation}
\langle:A(\omega)B(\omega+\delta\omega:\rangle:=-i\lim_{\delta\omega\rightarrow0}[\delta\omega\langle A(\omega)B(\omega+\delta\omega)\rangle].
\end{equation}
Here we introduce a notation $\langle:AB:\rangle$ for the simplicity
of the limitation above. With this notation, From Eq.\,(\ref{eq:fluc})
we obtain that the fluctuation of the above random forces gives
\begin{align}
\langle:\tilde{\eta}_{x}^{\dagger}(\omega)\tilde{\eta}_{y}(\omega+\delta\omega):\rangle & =\Gamma_{x}(\omega)f_{x}(\omega-\mu_{x})\delta_{xy},\nonumber \\
\langle:\tilde{\eta}_{x}(\omega)\tilde{\eta}_{y}^{\dagger}(\omega+\delta\omega):\rangle & =\Gamma_{x}(\omega)f_{x}(\mu_{x}-\omega)\delta_{xy}.
\end{align}

Further, recall that
\begin{align*}
\tilde{\mathbf{d}}(\omega) & =\tilde{\mathbf{G}}(\omega)[\mathbf{d}(0)-i\tilde{\boldsymbol{\eta}}(\omega)],\\
\tilde{\Psi}_{x}(\omega) & =\tilde{\eta}_{x}(\omega)-i\tilde{D}_{x}(\omega)\tilde{d}_{x}(\omega),
\end{align*}
 we have
\begin{align*}
\langle:\tilde{d}_{x}^{\dagger} & \tilde{\Psi}_{x}:\rangle=\langle:i\sum_{j}\tilde{\boldsymbol{\eta}}_{j}^{\dagger}\mathbf{G}_{jx}^{\dagger}\tilde{\eta}_{x}-i\sum_{i,j}\tilde{\boldsymbol{\eta}}_{j}^{\dagger}\mathbf{G}_{jx}^{\dagger}D_{x}\mathbf{G}_{xi}\tilde{\boldsymbol{\eta}}_{i}\\
+ & \sum_{j}\hat{d}_{j}^{\dagger}(0)\tilde{\mathbf{G}}_{jx}^{\dagger}(\tilde{\eta}_{x}-iD_{x}\tilde{d}_{x})+\sum_{i,j}\tilde{\boldsymbol{\eta}}_{j}^{\dagger}\tilde{\mathbf{G}}_{jx}^{\dagger}D_{x}\tilde{\mathbf{G}}_{xi}\hat{d}_{i}(0):\rangle.
\end{align*}
 The last two terms containing information from the initial state
vanish to zero after long time evolution. This can be also verified
by calculating the limitation. Some straightforward calculation shows
that the steady current Eq.\,(\ref{eq:steady-current}) becomes
\begin{align}
\overline{I}_{x} & (t\rightarrow\infty)=-\frac{ie}{\hbar}\int\frac{d\nu}{2\pi}\langle:\tilde{d}_{x}^{\dagger}\tilde{\Psi}_{x}-\Psi_{x}^{\dagger}d_{x}:\rangle\\
 & =\frac{e}{\hbar}\int\frac{d\nu}{2\pi}\langle:\tilde{\eta}_{x}^{\dagger}[\tilde{\mathbf{G}}^{\dagger}\tilde{\boldsymbol{\Gamma}}\tilde{\mathbf{G}}]_{xx}\tilde{\eta}_{x}-\sum_{i}\tilde{\boldsymbol{\eta}}_{i}^{\dagger}\tilde{\mathbf{G}}_{ix}^{\dagger}\Gamma_{x}\tilde{\mathbf{G}}_{xi}\tilde{\boldsymbol{\eta}}_{i}:\rangle.\nonumber 
\end{align}
Here we used the following relation,
\begin{align}
\tilde{\mathbf{G}}+\tilde{\mathbf{G}}^{\dagger} & =\tilde{\mathbf{G}}\big[i(\omega-\mathbf{H}-i\tilde{\mathbf{D}}^{\dagger})-i(\omega-\mathbf{H}+i\tilde{\mathbf{D}})\big]\tilde{\mathbf{G}}^{\dagger}\nonumber \\
 & =\tilde{\mathbf{G}}\tilde{\boldsymbol{\Gamma}}\tilde{\mathbf{G}}^{\dagger}=\tilde{\mathbf{G}}^{\dagger}\tilde{\boldsymbol{\Gamma}}\tilde{\mathbf{G}},
\end{align}
where $\tilde{\boldsymbol{\Gamma}}=\tilde{\mathbf{D}}+\tilde{\mathbf{D}}^{\dagger}$
is the dissipation matrix. Finally we can write down $\overline{I}_{x}$
in sum of components as
\begin{align}
\overline{I}_{x}(t\rightarrow\infty)= & \frac{e}{\hbar}\int\frac{d\nu}{2\pi}\tilde{\mathbf{G}}_{yx}^{\dagger}\Gamma_{x}\tilde{\mathbf{G}}_{xy}\Gamma_{y}(f_{x}-f_{y})\\
 & +\tilde{\mathbf{G}}_{x+N,x}^{\dagger}\Gamma_{x}\tilde{\mathbf{G}}_{x,x+N}\Gamma_{x}(f_{x}-\overline{f}_{x})\nonumber \\
 & +\tilde{\mathbf{G}}_{y+N,x}^{\dagger}\Gamma_{y}\tilde{\mathbf{G}}_{x,y+N}\Gamma_{x}(f_{x}-\overline{f}_{y}),\nonumber 
\end{align}
where we denote $f_{x}:=f_{x}(\nu-\mu_{x})$ and $\overline{f}_{x}:=f_{x}(\mu_{x}-\nu)=1-f_{x}$.

\end{document}